\documentclass{article}

\usepackage{authblk}
\usepackage[font=small]{caption}
\usepackage{booktabs}
\usepackage{graphicx}
\usepackage{caption}
\usepackage{lineno}
\usepackage{chngcntr}
\usepackage{mathtools}
\usepackage{comment}
\newtagform{brackets}{[}{]}
\usetagform{brackets}
\usepackage[toc,page,titletoc]{appendix}
\usepackage[capitalise,nameinlink]{cleveref}

\let\citeleft=(
\let\citeright=)

\begin{document}

\pdfinfo{
   /Author (AUTHORS)
   /Title (Free-breathing motion compensated 4D (3D+respiration) T2-weighted turbo spin-echo MRI for body imaging)
}

\title{\vspace{-2cm}Free-breathing motion compensated 4D (3D+respiration) T2-weighted turbo spin-echo MRI for body imaging}

\author[1,2]{T. Bruijnen}
\author[1]{T. Schakel}
\author[1,2,3]{O. Akdag}
\author[1,2,3]{C.V.M. Bruel}
\author[1,2,3]{J.J.W. Lagendijk}
\author[1,2]{C.A.T. van den Berg}
\author[1,4]{R.H.N. Tijssen}

\affil[1]{Department of Radiotherapy, University Medical Center Utrecht, Utrecht, the Netherlands}
\affil[2]{Computational Imaging Group for MRI diagnostics and therapy, Centre for Image Sciences, University Medical Center Utrecht, Utrecht, the Netherlands}
\affil[3]{Department of Biomedical Engineering, Eindhoven University of Technology, Eindhoven, the Netherlands}
\affil[4]{Department of Radiation Oncology, Catharina Hospital, Eindhoven, the Netherlands}

\maketitle

\vfill
\noindent
\textit{Running head:} Free-breathing motion compensated $T_2$-w-TSE

\noindent
\textit{Address correspondence to:} \\
  Tom Bruijnen, Department of Radiotherapy, University Medical Center Utrecht, Utrecht, the Netherlands \\
  t.bruijnen@umcutrecht.nl

\noindent
This work is part of the research programme HTSM with project number 15354, which is (partly) financed by the Netherlands Organisation for Scientific Research (NWO) and Philips Healthcare.


\noindent

\noindent
Submitted to \textit{Magnetic Resonance in Medicine} as a Full Paper.

\clearpage
\section*{Abstract}

\noindent
\textbf{Purpose}: To develop and evaluate a free-breathing respiratory motion compensated 4D (3D+respiration) $T_2$-weighted turbo spin echo sequence with application to radiology and MR-guided radiotherapy.

\noindent 
\textbf{Methods}: k-space data are continuously acquired using a rewound Cartesian acquisition with spiral profile ordering (rCASPR) to provide matching contrast to the conventional linear phase encode ordering and to sort data into multiple respiratory phases. Low-resolution respiratory-correlated 4D images were reconstructed with compressed sensing and used to estimate non-rigid deformation vector fields, which were subsequently used for a motion compensated image reconstruction.

\noindent rCASPR sampling was compared to linear and CASPR sampling in terms of  point-spread-function (PSF) and image contrast with \textit{in silico}, phantom and \textit{in vivo} experiments. Reconstruction parameters for low-resolution 4D-MRI (spatial resolution and temporal regularization) were determined using a grid search. The proposed motion compensated rCASPR was evaluated in eight healthy volunteers and compared to free-breathing scans with linear sampling. Image quality was compared based on visual inspection and quantitatively by means of the gradient entropy. 

\noindent
\textbf{Results}: rCASPR provided a superior PSF (similar in ky and narrower in kz) and showed no considerable differences in images contrast compared to linear sampling. The optimal 4D-MRI reconstruction parameters were spatial resolution=4.5 mm3 (3x reduction) and $\lambda_t = 1 \cdot 10^{-4}$. The groupwise average gradient entropy  was $22.31 \pm 0.07$ for linear, $22.20\pm 0.09$ for rCASPR, $22.14\pm0.10$ for soft-gated rCASPR and $22.02\pm0.11$ for motion compensated rCASPR.

\noindent
\textbf{Conclusion}: The proposed motion compensated rCASPR enables high quality free-breathing T2-TSE with minimal changes in image contrast and scan time. The proposed method therefore enables direct transfer of clinically used 3D TSE sequences to free-breathing.

\noindent
\textbf{Keywords}: Motion correction, turbo spin-echo, fast spin-echo, motion compensated image reconstruction, compressed sensing,
\clearpage

\section{Introduction}
\label{sec:introduction}
Respiratory motion during Magnetic Resonance Imaging (\textbf{MRI}) has been a long standing problem that leads to considerable reductions in image quality \cite{Zaitsev2015}. The solutions traditionally proposed in the radiology workflow are breathhold or respiratory triggered scans, which are effective at reducing motion artefacts at the cost of limited spatial resolution, increased scan time and reduced patient comfort. The solutions traditionally proposed in the MR-guided radiotherapy workflow are respiratory correlated 4D-MRI scans \cite{Stemkens2018}, which sort the data into multiple motion states based on a respiratory surrogate signal. These 4D-MRI scans simultaneously reduce motion artefacts and quantify the respiratory motion, but require considerably longer scan times and often show reduced image quality compared to breathhold and gated scans. This reduction in image quality is one of the main reasons why 4D-MRI methods are not widely adopted in clinical exams. Recent advances using motion robust sampling trajectories~\cite{Winkelmann2007,Block2014}, self-navigation~\cite{Spincemaille2011,Zhang2016}, compressed sensing image reconstruction \cite{Feng2014,Feng2016} and motion compensated image reconstruction \cite{Batchelor2005,Buerger2013} have improved image quality considerably, but are often limited to $T_1$-weighted ($T_1$-w) gradient echo (\textbf{GRE}), while many clinical applications require $T_2$-weighted ($T_2$-w) turbo spin-echo (\textbf{TSE}) scans \cite{Mugler2014,Freedman2018}. \noindent $T_{2}$-w TSE has clinical utility in multiple abdominal applications, ranging from the diagnostics of focal hepatic masses \cite{Watanabe2007} and biliary disorders \cite{Arizono2008} to contouring of pancreatic tumours for radiation therapy \cite{Heerkens2017}. Therefore, there is a clear need for a motion robust 4D (3D+respiration) free-breathing sequence that is able to provide similar image contrast as conventional $T_2$-w TSE exams.

\noindent $T_{2}$-w TSE scans are typically acquired using 2D multi-slice methods, which offer lower spatial resolution in the slice direction (compared to 3D). The low spatial resolution in the slice direction impedes retrospective reformatting into freely selectable view orientations, which is diagnostically favorable for specific applications \cite{Albiin2012}. Motion robust 3D $T_2$-w TSE scans in free-breathing are currently uncommon in clinical applications and have sparked little interest in research. One of the reason is the inherent challenges accompanied with high quality 4D (3D+respiration) $T_2$-w TSE. The main challenge is the reduced flexibility in selection of k-space trajectory, because non-Cartesian schemes (e.g. radial) repetitively sample the k-space center and therefore impact the $T_2$-weighting. These differences in $T_2$-weighting lead to an undefined image contrast and, as pointed out by Benkert et al. \cite{Benkert2018}, can only be circumvented by using k-space filtering \cite{Song2000} techniques or model-based reconstructions \cite{Block2009}. Alternatively, Benkert et al. proposed a stack-of-stars trajectory to place the echo train along the Cartesian sampled direction ($k_z$) \cite{Benkert2018}, which directly couples the echo train length to the number of partitions in the $k_z$ direction. The second challenge is that the acquisition times of 3D $T_2$-w TSE are longer than for GRE (2-3 times), which  requires aggressive undersampling to reduce scan times to clinically acceptable levels and subsequently requires efficient image reconstruction algorithms to cope with the highly undersampled data \cite{Rank2017}.

\noindent In this work we further develop motion robust free-breathing 4D (3D+respiration) $T_2$-w TSE imaging in order to improve the image quality to make it comparable to respiratory triggered or breathhold examinations. First, we propose a novel k-space trajectory called a rewound spiral acquisition with spiral profile ordering (rCASPR), which is an extension of the regular CASPR trajectory \cite{Prieto2015}. A kin to CASPR, rCASPR provides self-navigation, golden angle profile ordering and inherent motion robustness due to variable density sampling \cite{Greer2019}. In addition, rCASPR increases the maximum viable echo train length to better accommodate $T_2$-contrasts for TSE imaging, which allows direct matching of the $T_2$ contrast to the conventional linear sampled 3D TSE scans. Second, we perform a motion compensated image reconstruction \cite{Batchelor2005} similar to the work of Kolbitsch et al. \cite{Kolbitsch2018}, in which the motion fields are estimated from a high resolution respiratory-correlated 4D-MRI. Instead of using the high resolution 4D-MRI, we propose to estimate the motion fields from a low resolution 4D-MRI, which according to previous works does not significantly reduce the quality of the deformation vector fields \cite{Glitzner2015a} \cite{Huttinga2020}. The reduction in spatial resolution is more aggressive than the approaches of previous works that used only a 1.5x reduced spatial resolution \cite{Zhu2020} or high spatial regularization \cite{Cruz2016} and considerably speeds-up the computation of the 4D-MRI. 

\noindent The proposed free-breathing 4D $T_2$-w TSE implementation is investigated in eight healthy volunteers. First, we assess the point-spread-function (PSF) and similarity in image contrast between rCASPR and conventional linear sampling. Second, we determine the optimal hyperparameters (spatial resolution and regularization parameter) for 4D-MRI based motion estimation. The quality of the motion fields are evaluated by using them in the motion compensated image reconstruction and subsequently evaluating the reconstructed images using the gradient entropy metric \cite{McGee2000}. Third, the general image quality of motion compensated rCASPR is compared against conventional linear sampling, rCASPR without motion correction and rCASPR with soft-gating \cite{Johnson2012,Cheng2015} using the gradient entropy metric \cite{McGee2000}. 

\section{Methods}
\label{sec:methods}

\subsection{MRI acquisition and preprocessing}
\begin{figure}[!ht]
\begin{center}
   \includegraphics[width=16cm]{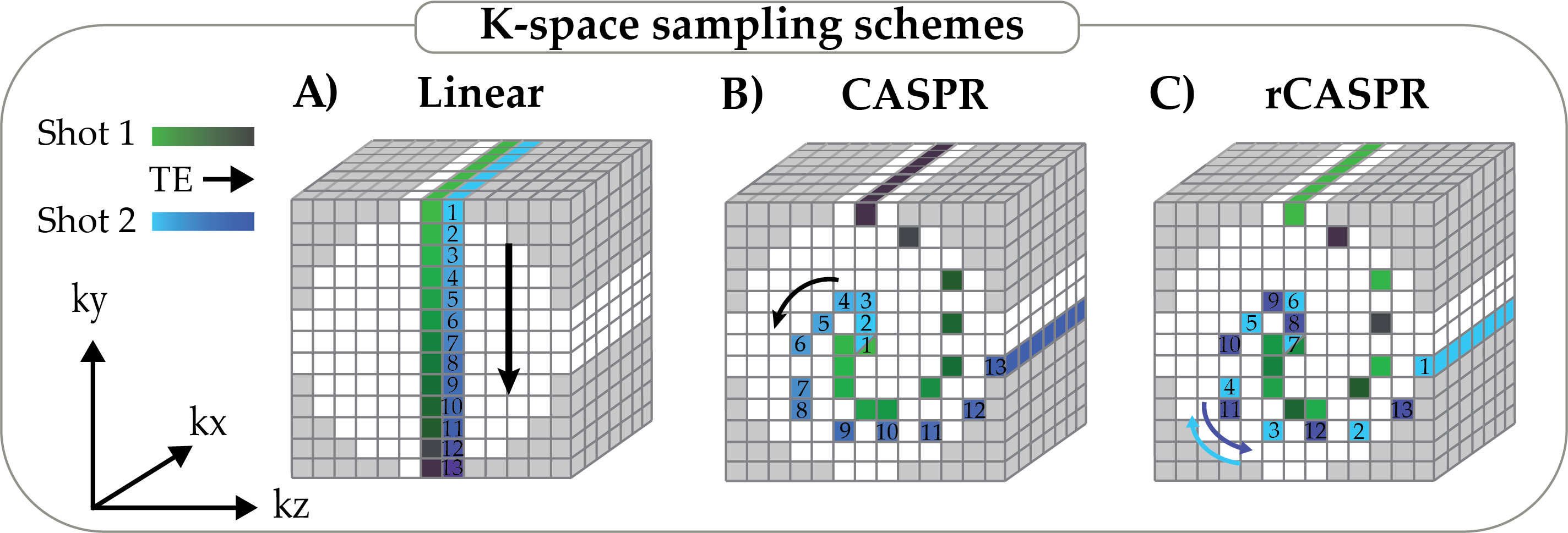}
 \vspace{-5pt}
\caption[]{\label{fig:fig1}  \textbf{K-space sampling schemes for turbo spin-echo imaging. A)} The linear scheme samples phase encodes along linear lines, which is widely used in clinical TSE sequences. \textbf{B)} The CASPR scheme samples phase encodes along spiral interleaves starting in the center and moving to the periphery (spiral-out). The order of phase encodes is indicated with the numbers, starting from 1 and moving to 13. Subsequent spirals are rotated with the golden ratio 137.5$^{\circ}$. \textbf{C)} The rCASPR scheme samples phase encodes along spiral interleaves, starting on the periphery, moving inwards and moving outwards again, i.e. spiral-in-out (as indicated by the numbers).  \textit{CASPR = Cartesian acquisition with spiral profile ordering, rCASPR = rewound Cartesian acquisition with spiral profile ordering, TSE = turbo spin-echo.}  }\vspace{-5pt}
\end{center}
\end{figure}

\noindent All subjects were scanned on a 1.5T radiotherapy MR system (Philips, Ingenia, Best, the Netherlands) using a 12 channel posterior and 16 channel anterior receive coil. This  study  was  approved  by  the  local  institutional  review  board and informed consent was obtained from all the participants. Eight volunteers were scanned with a TSE sequence with variable refocusing flip angle train to maximize the shot length for the desired $T_2$ contrast. Relevant sequence parameters for both acquisitions are shown in Table~\ref{tbl:1}. All scans were acquired two times, once with the standard linear scheme and once with the proposed golden angle rCASPR scheme (Fig~\ref{fig:fig1}). 
rCASPR samples phase encodes on rotated spiral interleaves, including the $k_{y,z} = (0,0)$ for each interleave, similar to CASPR. However, rCASPR starts at the periphery and samples half of the phase encodes going inwards and samples the other half of the phase encodes going outwards (Fig~\ref{fig:fig1}C). In other words, CASPR samples a spiral-out while rCASPR samples a spiral-in-out. The rCASPR phase encodes are selected with a nearest neighbour interpolation from an analytical Archimedean spiral with one revolution per shot. The rCASPR sampling scheme ensures that no duplicate phase encodes are acquired within one spiral interleave. Between the different spiral interleaves, the $k_{y,z} = (0,0)$ phase encode is repeated in order to enable self-navigation and motion estimation. Contrast control to specific effective echo times can be achieved by shifting the entire spiral interleave. This will shift the acquisition of the $k_{y,z} = (0,0)$ phase encodes earlier/later in the TSE train. The shifted phase encodes are added to the start or end of the TSE train, depending on the direction of the shift.  More details on the sampling scheme are provided on github \cite{Bruijnen2021}.

\begin{table}[!ht]
\renewcommand{\tabcolsep}{0.1cm}
 \caption{\textbf{Scanner and sequence parameters of the volunteer imaging experiments.}}
\begin{center}
           \small{
          \begin{tabular}{l l }
            \toprule
           \multicolumn{2}{l}{\textbf{Sequence settings}} \\ 
            & \textbf{$T_2$-TSE}  \\
            \midrule
            Field strength & 1.5T  \\
            Spatial resolution & 1.3 x 1.3 x 2.0 mm\textsuperscript{3}  \\
            Matrix size & 308 x 270 x 133 \\
            Field-of-view & 400 x 350 x 400 mm\textsuperscript{3} \\
            Repetition time & 1000 ms \\
            Effective echo time & 230 ms\\
            Echo train length & 114  \\
            Echo spacing & 3.7 ms \\
            Readout bandwidth & 855 Hz/pixel \\
            Scan time & 248 s \\

  \bottomrule
 \end{tabular}
 }
\end{center}
 \label{tbl:1}
\end{table}

\begin{figure}[!ht]
\begin{center}
   \includegraphics[width=16cm]{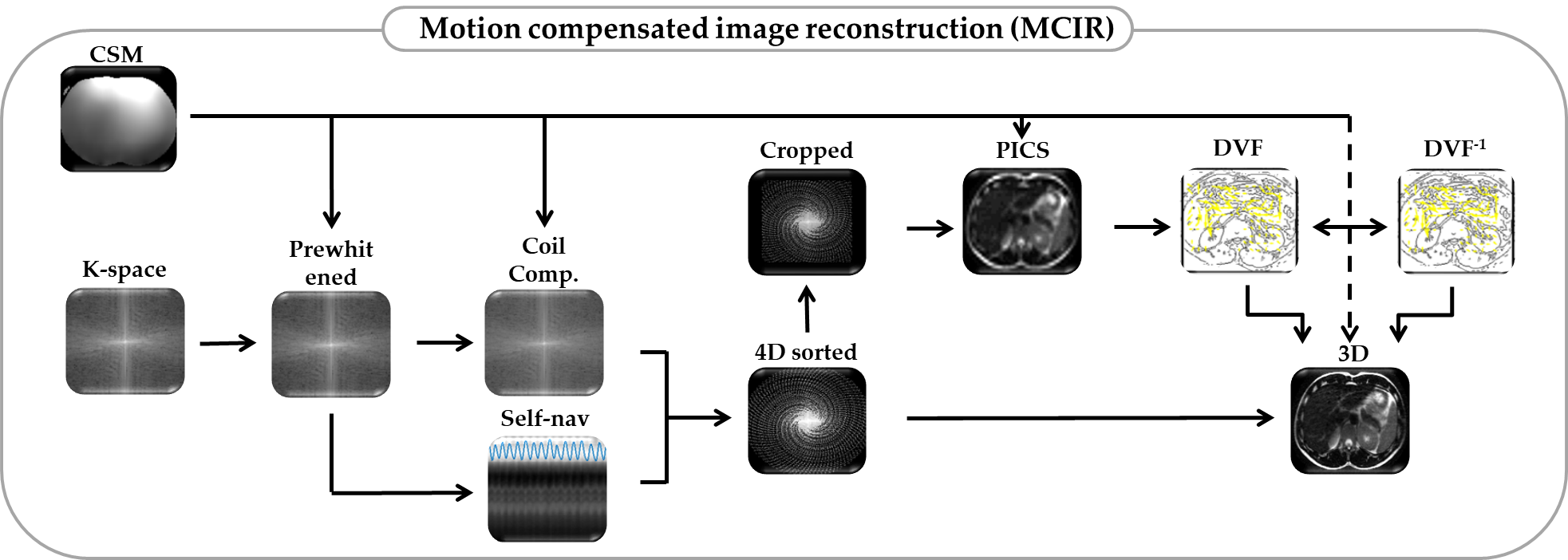}
 \vspace{-5pt}
\caption[]{\label{fig:fig2} \textbf{Overview of the proposed image reconstruction framework.} Input consists of measured k-space data and a pre-scan to estimate the CSM and noise levels. The k-space data is prewhitened and coil compressed using information from the pre-scan. In parallel, the self-navigation profile is estimated from the multi-coil $k_z$ projections. The k-space data is binned in respiratory phases and subsequently cropped to lower spatial resolution. The low resolution k-space is reconstructed using PICS and registered to obtain the DVFs. The DVFs are inverted (DVFs$^{-1}$) and interpolated to high resolution. The high resolution DVFs are used together with the high resolution binned k-space data for the motion compensated image reconstruction. \textit{CSM = coil sensitivity maps, PICS = parallel imaging and compressed sensing, DVF = deformation vector field}}\vspace{-5pt}
\end{center}
\end{figure}

\noindent Prior to the TSE sequences a reference scan, implemented by the vendor, was acquired to generate the coil sensitivity maps (\textbf{CSM}) \cite{Pruessmann1999}. K-space data were exported using Reconframe (Gyrotools, Zurich, CH) and were prewhitened and coil compressed to 16 virtual channels. The multi-channel $k_{x,y} = (0,0)$ projections were used to estimate the respiratory motion surrogate using the coil clustering method \cite{Zhang2016,Feng2016}. The motion surrogate was subsequently used to soft-gate or to reorder the data across respiratory phases using amplitude binning. Note that both the soft-gating and data reordering operate on complete spiral interleaves and not on partial spiral interleaves. All images were reconstructed offline in Matlab and a complete overview of the reconstruction framework is shown in Fig~\ref{fig:fig2}. 

\subsection{Respiratory-correlated 4D-MRI reconstruction}
Prior to image reconstruction the k-space data were cropped to lower spatial resolution to reduce the undersampling factor. The extent of reduction in spatial resolution will be addressed in subsection 2.8. The surrogate signal was used to reorder the k-space data across 8 respiratory phases using soft-gating with a Gaussian kernel that partially included data from neighbouring phases \cite{Johnson2012,Cheng2015,Han2017}. The soft-weights ($W_t$) were constrained to restrict the total number of samples to $\sum_{i=1}^{N_{spirals}} W_{t,i} = 1.5 \cdot \frac{N_{spirals}}{N_{phases}}$, with $N_{spirals}$ as the number of acquired spiral interleaves and $N_{phases}$ as the number of respiratory phases. Respiratory resolved images were reconstructed using parallel imaging and compressed sensing (\textbf{PICS}) with locally low-rank constraints \cite{Jiang2018} using the BART toolbox \cite{Uecker2015}:  

\begin{eqnarray}
x_t = \min\limits_{x_t} ||W_t(FSx_t - y_t)||_2 + \lambda_t ||x_t||_* 
\end{eqnarray}

\noindent With $W_t$ as the soft-weights for the different respiratory phases, $F$ as the Fast Fourier Transform, $S$ as the coil sensitivities, $x_t$ as the respiratory phases images, $y_t$ as the respiratory sorted k-space data, $\lambda_t$ as the regularization parameter and $||x_t||_*$ as the nuclear norm. 

\subsection{Motion estimation}
The low-resolution 4D images were interpolated (cubic) to the original high resolution matrix size. The respiratory phase images were registered to the exhale position to obtain the deformation vector fields (\textbf{DVF}) using Matlab's implementation of the diffeomorphic demons algorithm (\textit{imregdemons}) \cite{Thirion1998,Vercauteren2009}. The demons algorithm uses the mean squared error as an image similarity metric. The algorithm used three resolution levels with regularization parameter ($\alpha$) set to the voxel equivalent of 5 mm. Note that we also evaluated two other registration methods \cite{Zachiu2015a,Metz2011}, but found that the demons algorithm was better at handling residual aliasing artefacts in the images (not reported). Subsequently, the registration was repeated in the opposing direction to get an initial guess for the inverse DVFs (DVF\textsuperscript{-1}). The DVF and DVF\textsuperscript{-1} were subsequently post-processed to enforce inverse consistency as reported in \cite{Yang2008}.

\subsection{Motion compensated 3D-MRI reconstruction}
The final motion compensated image is reconstructed by incorporating the DVFs in the signal model: 

\begin{eqnarray}
    x = \min\limits_{x} ||FSU^Hx - y||_2 + \lambda_w ||\Psi x||_1 
\end{eqnarray}

\noindent With $U^H$ as the operator that warps the exhale image to the different respiratory phases and $x$ as the ideal exhale image, $y$ as the k-space data, $\lambda_w$ as the regularization parameter ($0.001$) and $\Psi$ as the wavelet transform as a sparse prior. The data consistency problem was solved using a modified version of the non-linear conjugate gradient algorithm provided by Feng et al. \cite{Feng2016} with gradient update step as proposed by Polak and Ribiere \cite{Polak1969}. More implementation details can be found in the publicly available code \cite{Bruijnen2021}.

\subsection{Soft-gated 3D-MRI reconstruction}
To compare the proposed motion compensated image reconstruction rCASPR with a retrospectively gated reconstruction, the data were reconstructed using soft-gating. All the k-space data were included to generate soft-weights ($W$) with an effective acceleration factor of 2, i.e. $\sum_{i=1}^{N_{spirals}} W_i = \frac{N_{spirals}}{2}$. The soft-weights were generated with the exhale position as a reference with a Gaussian weighting kernel \cite{Bruijnen2019a}. The weights were included in the PICS reconstruction using the BART toolbox \cite{Uecker2015}:

\begin{eqnarray}
x = \min\limits_{x} ||FSx - y||_2 + \lambda_w ||\Psi x||_1 
\end{eqnarray}

\noindent With $W$ as the soft-weights.

\subsection{Point spread function analysis: Linear vs. rCASPR}
The key concept of rCASPR sampling is to maintain the contrast of the conventional $T_2$ TSE sequences, which often use linear sampling. However, rCASPR samples high frequencies with a mix of short and long echo times, which inevitably leads to a small difference in image contrast and the corresponding $T_2$ point spread function. To investigate the difference in $T_2$ point spread function we setup a small \textit{in silico} experiment. The refocusing flip angle trains along with the subscribed phase encode scheme of the $in~vivo$ scans were extracted from the MR system. The flip angle patterns were used to simulate a TSE scan for a tissue type with $T_1 = 1000$ ms and $T_2 = 100$ ms using extended phase graphs \cite{Weigel2015}. The phase encode schemes were used to calculate echo time maps. The signal response and echo time maps were used to sample a single point in image space that exhibits $T_2$ decay along the echo train. The fully sampled k-space data was subsequently reconstructed using a fast Fourier Transform (with zero-padding) and the resulting 1D and 2D point spread functions (PSF) were compared between rCASPR and linear sampling. 

\subsection{Comparison of image contrast: Linear vs. rCASPR}
The rCASPR sampling scheme was introduced to facilitate 4D image reconstruction, while minimizing differences in image contrast. To assess these differences in image contrast two experiments were performed. First, a gel tube phantom with varying $T_1$ and $T_2$ values was measured with linear, CASPR and rCASPR scans. The mean value within the tubes was computed and compared using the normalized root mean squared error. Second, one dataset of a volunteer was analyzed to inspect differences in image contrast between linear and rCASPR reconstructions (no motion correction). The root mean square error between the images was calculated and line profiles were extracted from static anatomy and qualitatively compared.

\subsection{Optimal spatial resolution for 4D-MRI based motion estimation}
The rCASPR scheme samples the k-space with variable density and therefore has an increasing undersampling factor with respect to the radial coordinate. This property is exploited in the respiratory-correlated 4D-MRI reconstruction to reduce the undersampling factor by reducing the spatial resolution. The key idea here is that lowering the spatial resolution reduces residual aliasing artefacts and motion blurring in the PICS reconstruction, while not affecting the final motion estimation \cite{Glitzner2015a}. In addition, a reduction in spatial resolution considerably reduces the computational requirements. The reduction in spatial resolution could improve the motion estimation and ultimately lead to higher image quality for the motion compensated rCASPR. However, the quality of the motion estimation is difficult to quantify from the DVFs directly \cite{Zachiu2019}. Therefore we propose to measure the image quality of the motion compensated rCASPR reconstructions as a surrogate for the quality of the 4D-MRI based motion estimation. For the optimization a grid search was performed for one volunteer with spatial resolution $\in [1.5: 1.5: 7.5] \ mm^3$ and $\lambda_t \in [5.0 , 10.0, 20.0, 50.0, 100.0 ] \cdot 10^{-4}$. In total 25 4D-MRIs were used for motion compensated image reconstructions. These reconstructions were subsequently analyzed using the global gradient entropy (\textbf{GE}) to determine the optimal hyperparameters \cite{McGee2000}. These hyperparameters were expected to be representative for the complete group and were therefore used for all volunteer motion compensated image reconstructions.

\subsection{Comparison of image quality: Linear vs rCASPR}
The proposed motion compensated rCASPR was compared to linear, rCASPR and soft-gated rCASPR in eight healthy volunteers for $T_2$-w TSE. The image quality was compared by computing the (global) gradient entropy for all the reconstructed images \cite{McGee2000}. A reduction of the gradient entropy corresponds to an increase an image quality. Note that all reconstructions were performed with the same spatial regularization parameter ($\lambda_w$) such that the gradient entropy is reflective of the removal of motion artefacts.

\section{Results}
\label{sec:results}
\subsection{Point spread function analysis: Linear vs rCASPR}
\begin{figure}[!ht]
\begin{center}
  \includegraphics[width=16cm]{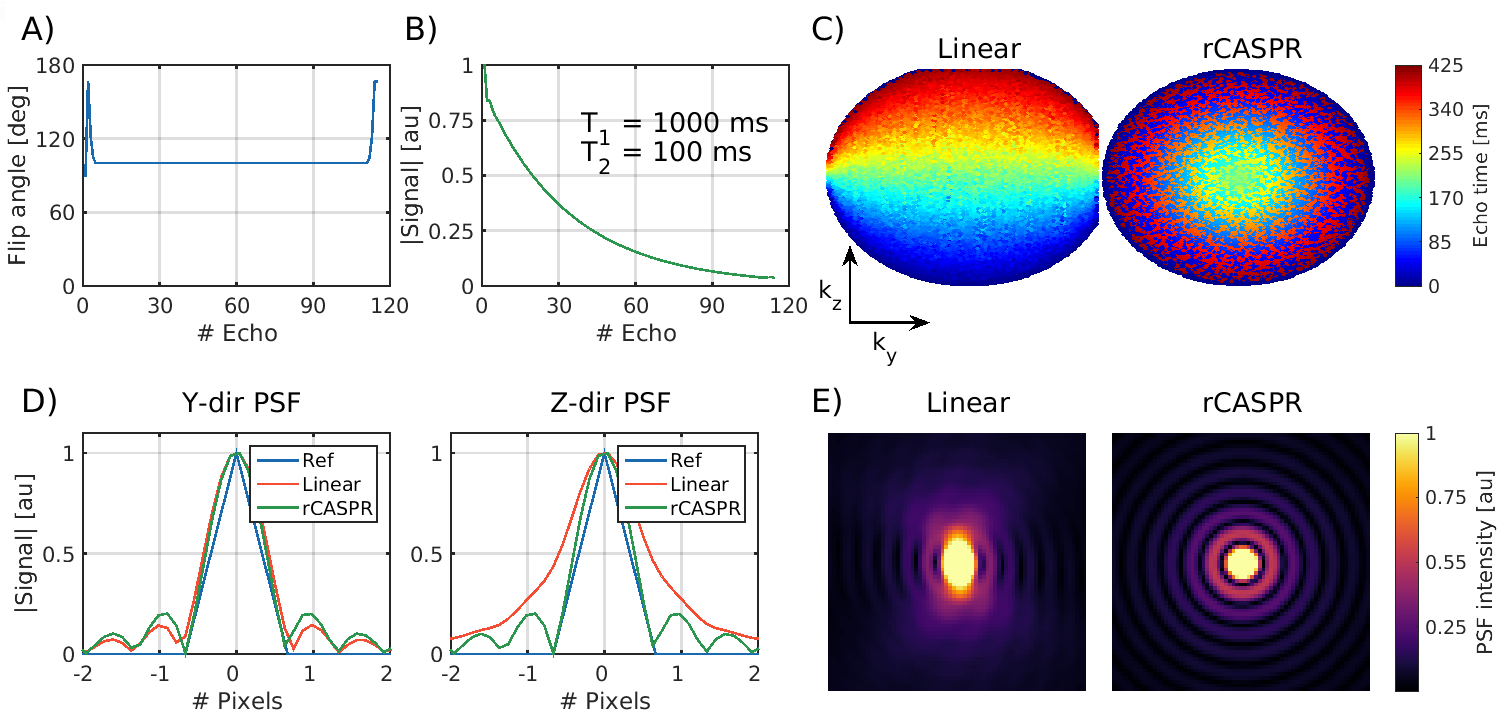}
 \vspace{-5pt}
\caption[]{\label{fig:fig3}\textbf{Point spread function analysis of linear and rCASPR sampling. A)} Refocusing angle train used for the experiments. \textbf{B)} Extended phase graph signal simulation for a spin with $T_1/T_2$ = 1000/100 ms. 
\textbf{C)} Echo time maps of all the sampled phase encodes. \textbf{D)} 1D point spread functions in the Y and Z directions. Note that the PSF of rCASPR is superior in the Z direction and similar in the Y direction compared to linear. \textbf{E)} 2D point spread functions for both linear and rCASPR.}\vspace{-5pt}
\end{center}
\end{figure}

Figure~\ref{fig:fig3} shows the result of the \textit{in silico} point spread function analysis. Figures-\ref{fig:fig3}A shows the flip angle train and B the corresponding $T_2$ decay that is used in combination with the echo time maps (Figure-\ref{fig:fig3}C) for linear and rCASPR sampling. Note that the echo time map of linear sampling shows that the phase encode scheme is optimized for acquisition efficiency and therefore does not sample straight columns. The echo time maps leads to the 1D and 2D point spread functions shown in Figure-\ref{fig:fig3}D+E. The point spread function is similar in the Y direction and narrower for rCASPR sampling in the Z direction. 

\subsection{Comparison of image contrast: Linear vs. rCASPR}
\begin{figure}[!ht]
\begin{center}
   \includegraphics[width=16cm]{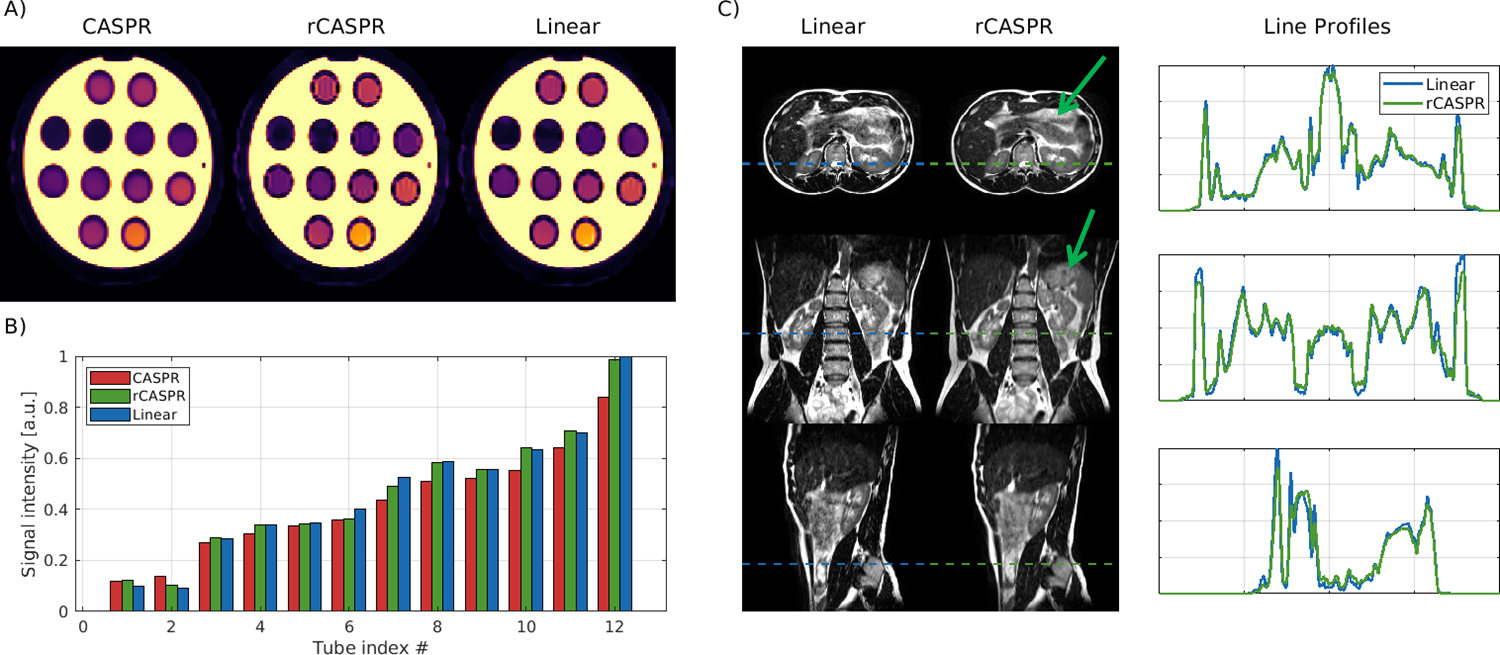}
 \vspace{-5pt}
\caption[]{\label{fig:fig4}\textbf{Image contrast comparison of linear vs rCASPR sampling schemes in $T_{2}$-TSE. 
A)} Gel tube phantom measurements of CASPR, rCASPR and Linear sampling schemes. \textbf{B)} Bar plots of the mean image intensities of the gel tubes for CASPR, rCASPR and Linear sampling. Note that rCASPR and Linear provide similar image intensities. \textbf{C)} Free-breathing abdominal 3D scans of a volunteer with Linear and rCASPR with identical scan parameters. Note that rCASPR shows reduced motion artefacts and slight blurring compared to the linear sampling scheme (green arrows). \textbf{D)} Line intensity profiles through the dashed lines from C).  \textit{Acronyms: rCASPR = rewound Cartesian acquisition with spiral profile ordering, TSE = turbo spin echo.} }\vspace{-5pt}
\end{center} 
\end{figure}

Figure~\ref{fig:fig4}A shows the phantom scan of the gel tube phantom with varying $T_1$ and $T_2$ values. Note that the image contrast of rCASPR is more similar to the linear scan then CASPR, which is also reflected in the normalized root mean squared error values that are 0.018 and 0.08 for rCASPR and CASPR respectively. These findings are also reflected in the mean signal intensities per tube as shown in the barplot in Figure~\ref{fig:fig4}B. Figure~\ref{fig:fig4}C shows \textit{in vivo} data of the linear and rCASPR sampling with RMSE = 0.051. The rCASPR reconstructions show reduced motion artefacts (green arrow) compared to linear sampling. The right side of the figure shows line intensity profiles, which are well aligned between the scans. 

\subsection{Optimal spatial resolution for 4D-MRI based motion estimation}
\begin{figure}[!ht]
\begin{center}
   \includegraphics[width=16cm]{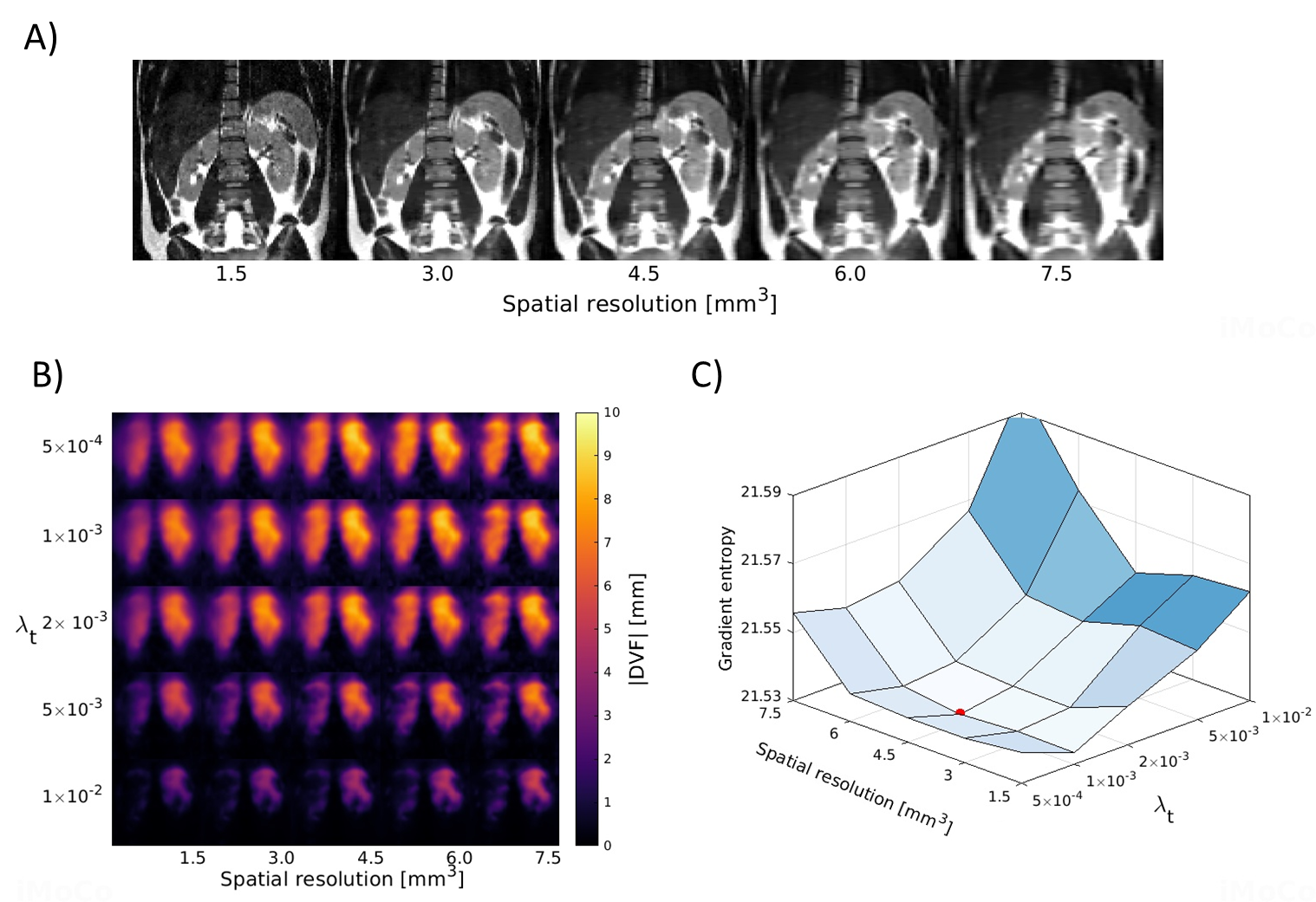}
 \vspace{-5pt}
\caption[]{\label{fig:fig5} \textbf{Hyperparameter grid search for optimal spatial resolution and temporal regularization ($\lambda_t$). A)} Top row shows a coronal slice reconstructed with five different spatial resolutions ($1.5-7.5mm^3$). \textbf{B)} The magnitude of the exhale-inhale motion fields derived from the 4D-MRIs with varying spatial resolution and temporal regularization. The investigated hyperparameters were in range: for spatial resolution $\in [1.5:1.5:7.5] \ mm^3$ and $\lambda_t \in [5.0 , 10.0, 20.0, 50.0, 100.0 ] \cdot 10^{-4}$. \textbf{C)} The optimization landscape that describes the gradient entropy for the motion compensated image reconstructions with the DVFs shown in panel B). Note that a lower gradient entropy corresponds with higher image quality and that the minimum of the function is indicated with red sphere. \textit{Acronyms: DVF = deformation vector field.} }\vspace{-5pt}
\end{center}
\end{figure}

Figure~\ref{fig:fig5}-A shows a coronal slice of the exhale position in the 4D-MRI for the different spatial resolution levels. The low resolution images show reduced aliasing artefacts, which are especially pronounced in the liver. Figure~\ref{fig:fig5}-B shows
the magnitude of the DVFs for the same coronal slice with varying temporal regularization and spatial resolution levels. There are two obvious trends within the data; the $|DVF|$ increases with reducing spatial resolution and that the $|DVF|$ also decreases with increasing regularization parameter $\lambda_t$. These trends seem to be stronger on the left side of the image, which corresponds with the anatomical position of the liver. The DVFs were subsequently used in the motion compensated image reconstruction. Figure~\ref{fig:fig5}-C shows the gradient entropy of the motion compensated image reconstructions for the varying varying regularization parameters and spatial resolution levels. The minimum within the parameter space is indicated with the red dot and corresponds with spatial resolution = $4.5 \  mm^3$ and $\lambda_t = 1\cdot10^{-3}$.

\begin{figure}[!ht]
\begin{center}
   \includegraphics[width=16.5cm]{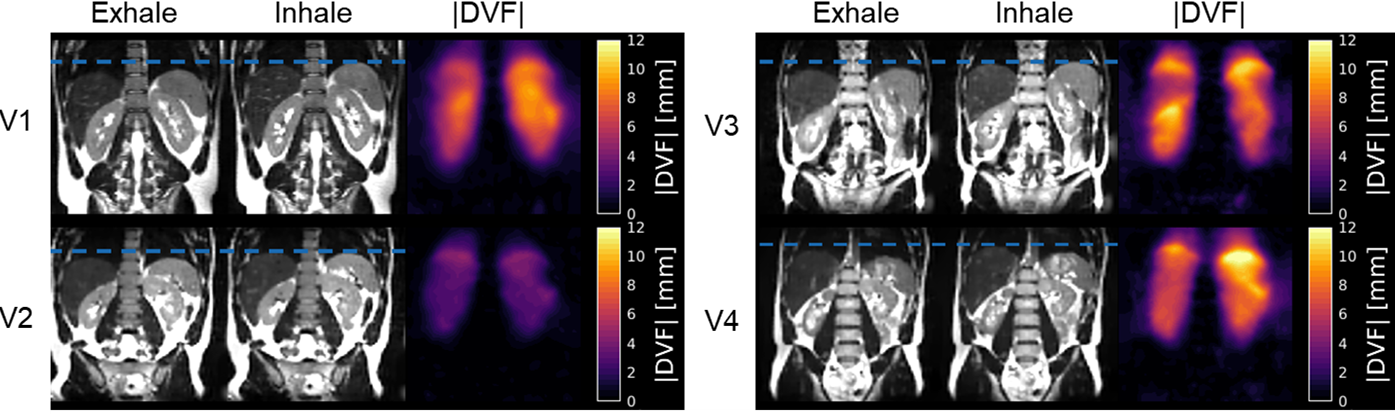}
 \vspace{-5pt}
\caption[]{\label{fig:fig6} \textbf{Respiratory correlated low-resolution 4D image reconstructions with the corresponding deformation vector fields.} The different volunteers are indicated with "Vx", where "x" refers to the volunteer index. The two grayscale images show the image reconstructed at the inhale and exhale position and the color coded images show the $|DVF|$ that describes the motion between these respiratory positions. Note that all volunteers show peak motion $\approx$ 1 cm with the exception of volunteer 2. The blue dashed line indicates the position of the liver/spleen in the exhale phase and aids in the visualization for the inhale phase. \textit{Acronyms: DVF = deformation vector field.}  }\vspace{-5pt}
\end{center}
\end{figure}

\noindent Figure~\ref{fig:fig6} shows the respiratory-correlated 4D-MRI reconstructions for four of the volunteers with the optimal hyperparameters described in the previous paragraph. The figure also shows the $|DVF|$ that describes the transformation between the exhale and inhale respiratory phase. Most volunteers show maximum displacement at the top of the liver larger than 1 cm, only volunteer 2 shows considerably smaller motion. As expected, almost no motion is visible in the static regions of the image (e.g. spine), indicating that residual aliasing artefacts do not affect the motion estimation. 

\subsection{Comparison of image quality: Linear vs rCASPR}
\begin{figure}[!ht]
\begin{center}
   \includegraphics[width=16cm]{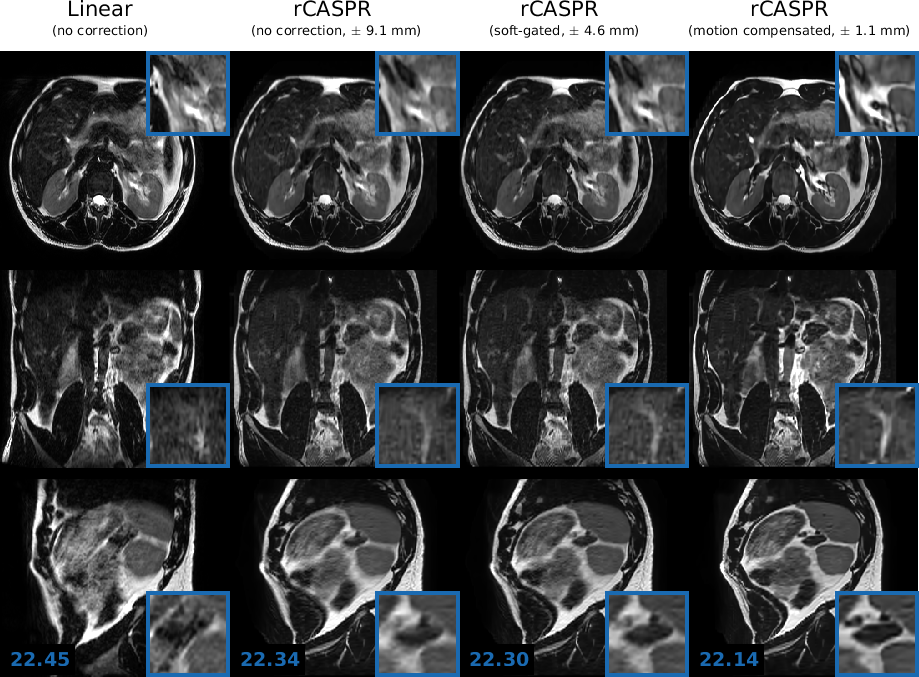}
 \vspace{-5pt}
\caption[]{\label{fig:fig8} \textbf{Volunteer 2: Comparison of image quality in in $T_2$-w free-breathing 3D turbo spin echo scans for linear, rCASPR, soft-gated rCASPR and motion compensated rCASPR.} For all the reconstructions the residual peak-to-peak motion was estimated using the 4D-MRI. For rCASPR the residual motion was $\pm 9.1 \ mm$, soft-gating rejected $50\% $ of the data such that the residual motion was $\pm 4.6 \ mm$ and motion compensation had residual intrabin motion of $\pm 1.1 \ mm$. The numbers in left bottom corner indicate the global gradient entropy. Blue boxes present zoomed regions. \textit{Acronyms: rCASPR = rewound cartesian acquisition with spiral profile ordering.}   }\vspace{-5pt}
\end{center}
\end{figure}

\begin{figure}[!ht]
\begin{center}
   \includegraphics[width=16cm]{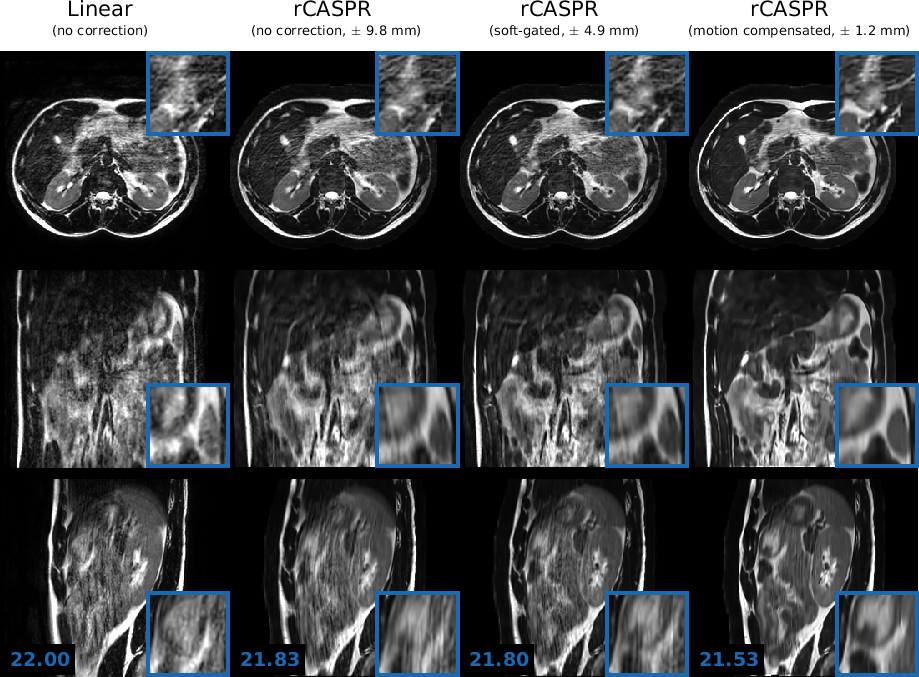}
 \vspace{-5pt}
\caption[]{\label{fig:fig7} \textbf{Volunteer 1: Comparison of image quality in in $T_2$-w free-breathing 3D turbo spin echo scans for linear, rCASPR, soft-gated rCASPR and motion compensated rCASPR.} For all the reconstructions the residual peak-to-peak motion was estimated using the 4D-MRI. For rCASPR the residual motion was $\pm 9.8 \ mm$, soft-gating rejected $50\% $ of the data such that the residual motion was $\pm 4.9 \ mm$ and motion compensation had residual intrabin motion of $\pm 1.2 \ mm$. The numbers in left bottom corner indicate the global gradient entropy. Blue boxes present zoomed regions. \textit{Acronyms: rCASPR = rewound cartesian acquisition with spiral profile ordering.}   }\vspace{-5pt}
\end{center}
\end{figure}

The comparison of image quality in two out of the eight volunteers are shown in Figures~\ref{fig:fig7}-\ref{fig:fig8}. The results of two other volunteers are shown in Supporting information I. Figure panels A-D show the linear, rCASPR, soft-gated rCASPR and motion compensated rCASPR reconstructions. Note that the soft-gated and motion compensated rCASPR are reconstructed at the exhale position. Zoom images of regions of interest are shown in blue boxes. Blue numbers at the left bottom display the gradient entropy calculated from the entire image. In general, image quality slightly increases when changing the sampling scheme from linear to rCASPR. The addition of soft-gating to rCASPR further increases the image quality. The largest improvement in image quality is observed when transitioning from no correction to the motion compensated image reconstruction. 

\noindent The groupwise average gradient entropy  was $22.31 \pm 0.07$ for linear, $22.20\pm 0.09$ for rCASPR, $22.14\pm0.10$ for soft-gated rCASPR and $22.02\pm0.11$ for motion compensated rCASPR. Supporting Information II shows a video of the motion compensated rCASPR reconstruction warped with the DVFs for all the volunteers.\newline

\section{Discussion}
\label{sec:discussion}
The key idea of this study was to develop a free-breathing 4D (3D+respiration) $T_2$-w TSE scan that matches the $T_2$ contrast of conventional linearly sampled 3D TSE scans while providing high quality motion corrected images. We proposed a sequence based on a novel golden angle rCASPR k-space trajectory that can maintain the desired $T_2$ contrast, while enabling self-navigation and respiratory correlated 4D-MRI. We proposed to reconstruct the 4D-MRI with reduced spatial resolution to obtain higher quality motion estimates with less computation time. The 4D-MRI reconstruction parameters (spatial resolution and temporal regularization) were optimized with a grid search guided by gradient entropy of the motion compensated image reconstruction. The optimal hyperparameters were spatial resolution = $4.5 \ mm^3$ (3x reduction) and $\lambda_t = 1 \cdot 10^{-3}$. The deformation vector fields were subsequently incorporated in the motion compensated rCASPR reconstruction, which was compared to linear, rCASPR and soft-gated rCASPR reconstructions. The image analysis indicated that motion compensated rCASPR provided the highest image quality in terms of the gradient entropy for all cases. 

\noindent An important aspect of this study that warrants discussion is the reliability and generalizability of the hyperparameters for the low resolution 4D-MRI reconstruction. The reliability is mainly dependent on the performance of the gradient entropy metric used as a surrogate for image quality. The gradient entropy was used in previous works on motion correction \cite{Cruz2016,Cheng2015,Odille2008a,Loktyushin2013} and McGee et al.\cite{McGee2000} reported that the gradient entropy corresponded best with observer scores out of 24 metrics in the context of motion correction. However, other image quality metrics may yield different optimal hyperparameters, which could potentially improve the image reconstructions. An alternative approach could be to use a metric directly on the motion fields, such as biomechanical quality assurance criteria \cite{Zachiu2019}. Note that the $4.5 mm^3$ found in this work correspond well with findings by Glitzner et al. \cite{Glitzner2015a}. The generalizability of the optimal hyperparameters is primarily dependent on the combination of the following variables: the acceleration factor, the image registration method and the scale of the physiological motion. For example, high acceleration will require reduced spatial resolution or high spatial regularization to mitigate aliasing artefacts or require an image registration algorithm that is resilient to aliasing artefacts \cite{Terpstra2020}. Therefore, the findings of optimal hyperparameters reported in this work are likely to be method-specific and therefore do not necessarily generalize to other combinations of acquisition and reconstruction methods. 

\noindent Overall, we believe that this work demonstrates the feasibility of free-breathing respiratory-correlated 4D (3D+respiration) $T_2$-w TSE. The combination of the rCASPR sampling scheme and motion estimation from low-resolution 4D-MRI allows high quality image reconstruction from a relatively short scan time ($<$ 5min). For radiology, the proposed motion compensated rCASPR could potentially replace respiratory triggered 3D $T_2$-w TSE scans, which could lead to simpler imaging workflows and improved robustness. Examples of these applications include: magnetic resonance urography, magnetic resonance cholangio-pancreatography and dynamic contrast enhanced imaging of the liver. However, future studies are required to compare the proposed free-breathing motion compensated rCASPR scan to prospectively respiratory triggered scans. For MR-guided radiotherapy, the motion compensated rCASPR provides a practical and efficient solution to quantify motion and ensure that the tumor is always covered by the radiation beam during free-breathing. In addition, the high quality motion compensated rCASPR images allow the definition of a highly accurate target contour, which is crucial for radiotherapy treatment planning. motion compensated rCASPR could also be used on a hybrid MRI-Linac to robustly assess day-to-day variations of the anatomy in free-breathing \cite{Lagendijk2008}. For PET/MRI, the motion compensated rCASPR could be used to conveniently maintain clinically used contrasts while concurrently quantifying the motion for the motion compensated PET reconstruction \cite{Rank2016,Kolbitsch2018}.

\section{Conclusion}
\noindent The proposed rCASPR sampling scheme, in combination with motion estimation from low-resolution 4D-MRI, enables high quality motion compensated image reconstruction. The proposed implementation enables direct transfer of contrast of 3D $T_2$-w TSE sequences to the free-breathing 4D (3D+respiration) $T_2$-w TSE counterparts with minimal changes in image contrast and scan time. Future studies are required to compare the image quality of the proposed method to respiratory triggered acquisitions, which are often the clinical standard.
\label{sec:conclusion}
\newpage

\section{Supporting Information}
\subsection{Supporting Information I}
\setcounter{figure}{0}
\renewcommand{\thefigure}{S\arabic{figure}}
\begin{figure}[!ht]
\begin{center}
\includegraphics[width=16cm]{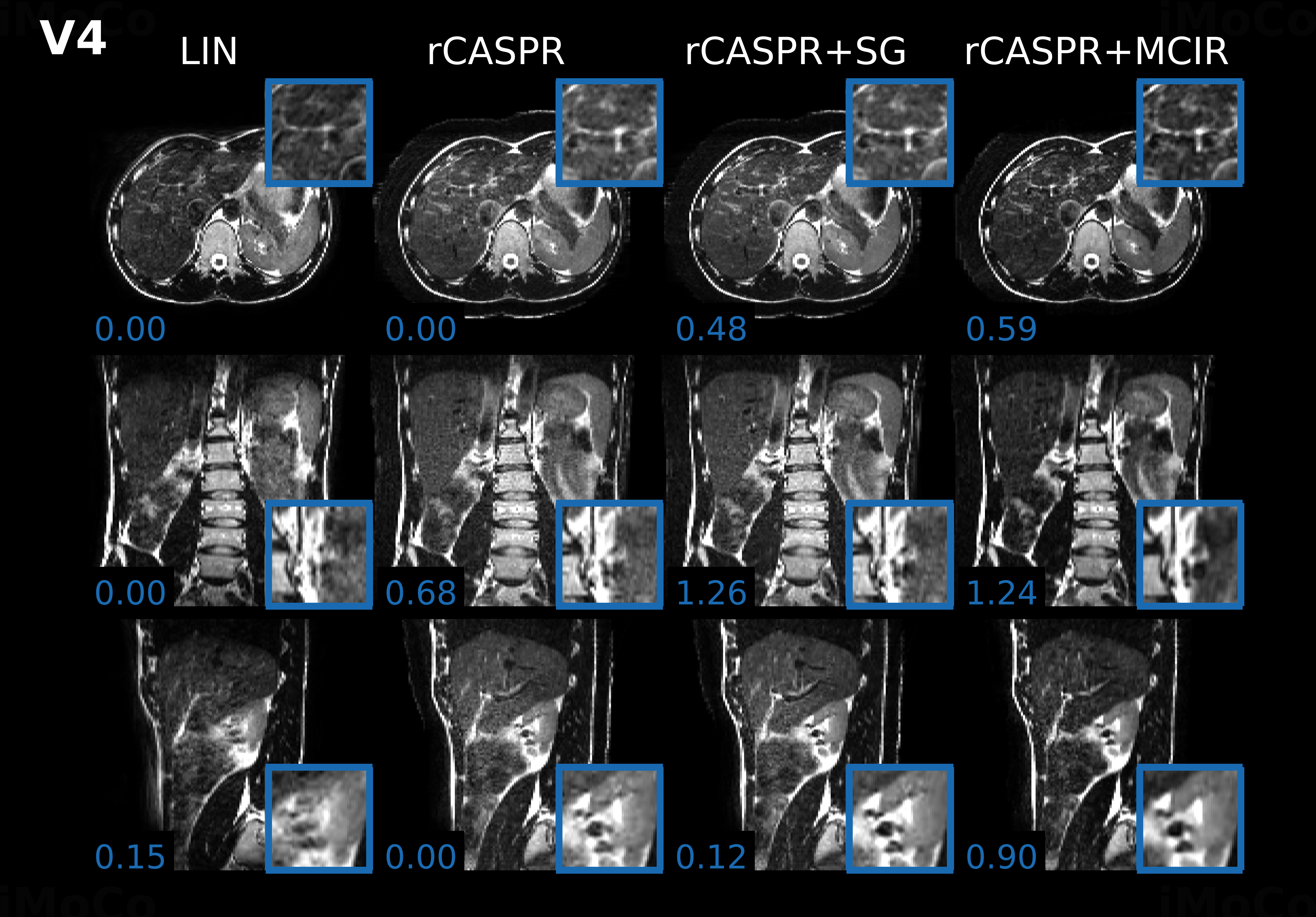}
 \vspace{-5pt}
\caption[]{\label{figS1:figS1} \textbf{Volunteer 3: TODO.}     }\vspace{-5pt}
\end{center}
\end{figure}

\subsection{Supporting Information I}

\begin{figure}[!ht]
\begin{center}
\includegraphics[width=16cm]{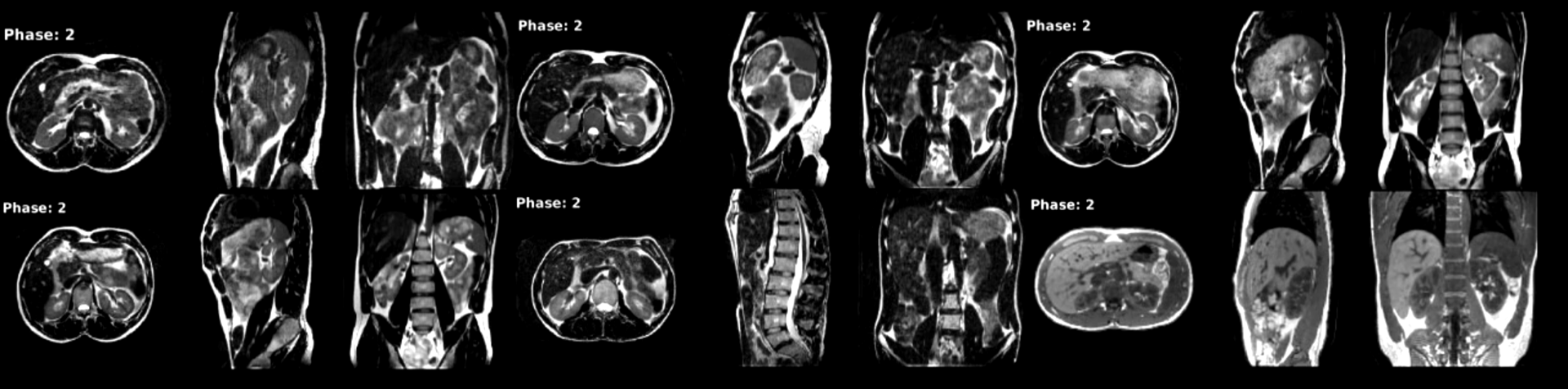}
 \vspace{-5pt}
\caption[]{\label{fig:figS2} \textbf{Motion compensated rCASPR images warped with the deformation vector fields used in the reconstruction.} Top row shows volunteer 1-3 and bottom row volunteer 4-6. Note that the 4D-MRI was reconstructed using 10 respiratory phases, but 20 are shown here for visualization purposes. Video version can be found: https://surfdrive.surf.nl/files/index.php/s/xmcAwLXRp51izcG }\vspace{-5pt}
\end{center}
\end{figure}

\end{document}